\documentclass[runningheads]{llncs}
\usepackage[T1]{fontenc}
\usepackage{graphicx}
\usepackage{multirow}
\usepackage{array}

\usepackage{eso-pic}

\begin{document}
\title{Quality Metrics for LLM-Generated Asset Administration Shells: A Perturbation-Based Evaluation Approach}
\titlerunning{Quality Metrics for LLM-Generated Asset Administration Shells}
%
\author{Janek Groß\inst{1}\orcidID{0000-0002-6306-711X} \and
Elena Zentgraf\inst{1}\orcidID{0009-0008-1031-2709} \and
Jens Heidrich\inst{1}\orcidID{0000-0001-6967-4722}}
\authorrunning{J. Groß et al.}
%
\institute{University of Applied Sciences Mainz, Mainz, Rhineland-Palatinate, Germany}
\maketitle              
\begin{abstract}
The rapid digital transformation of manufacturing, often referred to as Industry 4.0, relies on seamless interoperability between physical and software assets. A central enabler is the Asset Administration Shell (AAS), a standardized digital representation of such assets. Recent advances in large language models (LLMs) enable the generation of AAS submodels from unstructured sources such as product datasheets but raise challenges for quality assurance. In particular, unexpected errors, the lack of ground truth references, and the absence of standardized quality metrics hinder reliable adoption.
In this work, we evaluate quality metrics for AI-generated AAS using a perturbation-based evaluation framework. By systematically degrading AAS generation along multiple dimensions, we assess how well different metrics reflect quality changes. Based on a dataset of 200 products from multiple manufacturers, we generate 6,400 AAS instances using GPT-4o-mini, Qwen3, and DeepSeek-R1.
Our results show that metrics based on exact matching of property names and similarity-based soft matching of property values, in particular value-based recall and name-based F1 score, provide the most reliable indicators of quality degradation. Furthermore, we quantify the impact of different perturbation types and analyze differences across model families and product segments. These findings support the selection of suitable metrics, the tuning of LLM-based pipelines, and the integration of AI-generated AAS into industrial applications.

\end{abstract}
\AddToShipoutPictureFG*{%
  \AtPageUpperLeft{%
    \raisebox{-1.5cm}{%
      \hspace{2cm}%
      \parbox{0.8\textwidth}{%
        \itshape
    Preprint. The final publication is available in the Proceedings of the 52nd Euromicro Conference on Software Engineering and Advanced Applications (SEAA 2026), Springer.
      }%
    }%
  }%
}

\section{Introduction}
\begin{figure}
\includegraphics[width=\textwidth]{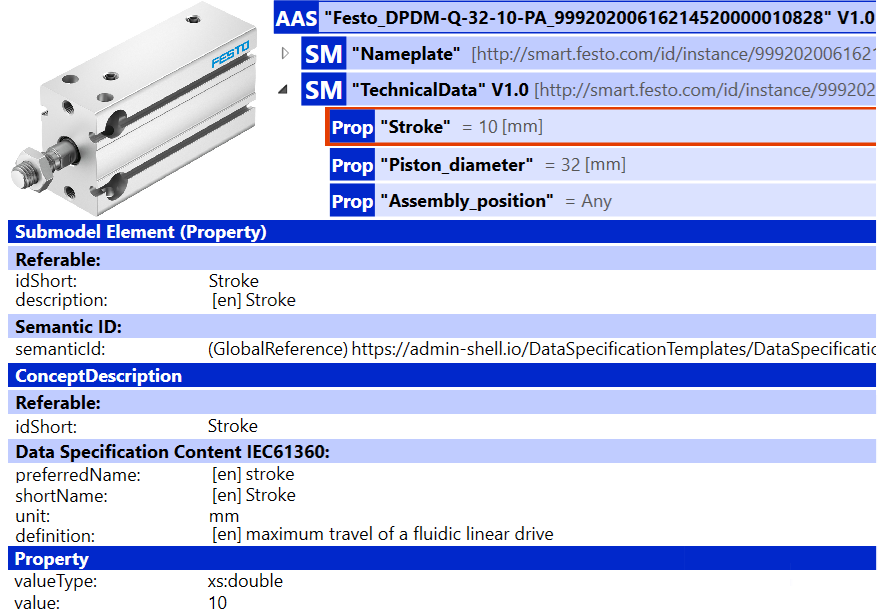}
\caption{Example AAS \cite{adminshell_samples} with three technical properties and property metadata.} 
\label{fig0}
\end{figure}

Asset Administration Shells (AAS) \cite{IndustrialDigitalTwinAssociation.2023,IndustrialDigitalTwinAssociation.2023b} provide standardized digital representations of hardware and software assets (see Fig.~\ref{fig0}). They define common semantics for properties and services and enable interoperable digital twins in Industry 4.0. As such, a large number of AAS must be created to cover the variety of assets used in modern manufacturing. 

Typically, manufacturers generate AAS from internal product data. However, if company-foreign or legacy products need to be modeled, manual extraction of properties from datasheets or technical documentation is required. This process is time-consuming, error-prone, and requires domain expertise.

Information extraction (IE) is the task to derive structured, machine-readable information from unstructured sources such as tables or natural-language text \cite{Han.2023,Deng.2024}. In the context of AAS creation, IE enables systematic extraction of product properties from technical documents. Traditionally, IE systems required extensive domain-specific annotation and complex pipelines. Recent advances in large language models (LLMs), particularly instruction tuning and function calling, have significantly lowered this barrier. LLMs now enable the construction of IE pipelines without supervised training, leading to emerging LLM-based AAS generation tools \cite{Xia.2024,Vogel.2025,Kaya.2025}. While these tools are promising, they remain unreliable and often produce unexpected results, raising critical questions about output validity and quality.

To enable trustworthy use of LLM-based tools for AAS generation, suitable quality metrics are necessary. Such metrics serve both the evaluation and optimization of generation tools including choice of model, prompting strategies and other hyperparameters. Defining suitable quality metrics, however, is non-trivial. AAS content is often non-unique: the same product can be described using different classification systems (e.g., ECLASS, ETIM, or company-specific schemes). Moreover, variations in naming, alternative units, or permissible structure complicate the application of standard IE metrics such as precision, recall, and F1 score. Furthermore, LLM-based generation performs multiple IE steps—such as entity recognition, normalization, and template mapping—in a single prompt-response cycle which prevents step-by-step evaluation. Together with the scarcity of labeled datasets, this underscores the need for tailored evaluation approaches.

In this work, we propose an empirical methodology for the evaluation of AAS quality metrics based on controlled perturbations of the generation process. By systematically degrading AAS instances, we analyze how well different metrics reflect changes in quality. Our research addresses the following questions:
(1) Which metrics best reflect AAS quality degradation under controlled perturbations?
(2) How do perturbation types and model characteristics affect AAS generation quality?
(3) How does AAS quality vary across manufacturers and product segments?

Our contributions are (i) a set of AAS-specific quality metrics, including full-reference and no-reference variants, (ii) a perturbation-based evaluation methodology for systematically assessing metric performance, and (iii) an empirical analysis of AAS quality across perturbation types, model families, manufacturers, and product segments. These contributions support the selection and benchmarking of AAS generation tools, enable automated quality assurance, and facilitate the integration of AI-assistants into industrial processes.

The remainder of this paper is structured as follows. Section \ref{background} examines relevant background and related work. In Section \ref{study_design}, the study design, including the evaluated metrics, the perturbation-based evaluation approach, and the experimental setup is described. Section \ref{results} presents empirical results. Section \ref{discussion} discusses the implications of the findings. Section \ref{threats} outlines threats to validity. Lastly, Section \ref{conclusion} concludes the paper and highlights directions for future work.

\section{Background and Related Work}
\label{background}
This section reviews relevant work on the evaluation of information extraction and LLM-based systems.

Traditional software quality models such as ISO/IEC 25010 \cite{ISO25010.2011,Wagner.2015} provide valuable frameworks for assessing software products based on characteristics such as functionality, reliability, and maintainability. However, these models primarily target conventional software and do not readily extend to structured artifacts such as AAS submodels. In contrast, our work focuses on the empirical evaluation of concrete quality metrics tailored to such AI-generated artifacts, rather than proposing an encompassing quality model.

The evaluation of generative models like LLMs poses a challenge due to the complexity of the output domain. It is common practice to evaluate LLMs on benchmark tasks with a unique, short answer to provide a general indication of model performance, but these benchmarks may not reflect performance in other, domain-specific tasks. For example, translation and summarization are commonly evaluated using metrics such as BLEU \cite{Papineni.2002} and ROUGE \cite{Lin.2004}, which rely on n-gram overlap between generated and reference texts. However, these metrics are not well suited for structured outputs, where fields are short and multiple semantically equivalent representations may exist.

Information extraction (IE) research is closely related to the AAS generation use case. IE systems typically combine tasks such as named entity recognition (NER) \cite{Li.2020}, relation extraction (RE) \cite{Zhao.2024}, and slot filling \cite{Witte.2022}, and are evaluated on labeled datasets using precision, recall, and F1 score. For AAS generation, however, labeled data for individual extraction steps is not available, and these IE steps (i.e. recognition of technical properties, value and unit normalization, mapping to AAS template) are often performed within a single LLM interaction. As a result, evaluation must be performed end-to-end and requires adaptations of classical metrics, such as soft matching for property names and values. These adaptations increasingly rely on AI-based methods like embedding-based similarity, to account for permissible variations and the absence of a unique correct output (e.g., color: gray vs. colour: grey).

Similar challenges arise in the evaluation of other multi-step LLM-based systems, such as retrieval-augmented generation (RAG) where output quality is difficult to assess directly. 
Recent work has therefore explored alternative evaluation strategies that closely align with our setting. These include indirect metrics based on downstream task performance and LLM-based judgments. These works also evaluate their methods through correlation with human judgments and test systems with intentionally varied retrieval quality.

For example, Sander and Dietz \cite{Sander.2021} assess RAG systems based on their ability to support follow-up questions, providing an indirect measure of output quality. Es et al. \cite{Es.2024} propose RAGAS, an automated framework that uses LLM prompting and embeddings to evaluate relevance and faithfulness. Building on that, Saad-Falcon et al. \cite{SaadFalcon.2024}  apply prediction-powered inference \cite{Angelopoulos.2023} to align automated evaluations with human judgments. To evaluate their approach, they simulate systems with controlled retrieval performance, which inspired the perturbation-based evaluation approach used in this work.

Overall, existing work highlights both the limitations of traditional evaluation methods for structured outputs and the need for more flexible, task-specific evaluation strategies. Building on insights from IE and RAG evaluation, our work introduces a framework for systematically benchmarking AAS quality metrics, even in the absence of high-quality labeled reference data.

\section{Study Design}
\label{study_design}
This study systematically evaluates quality metrics for AI-generated Asset Administration Shells (AAS) using a perturbation-based framework. We deliberately degrade AAS generation along multiple dimensions to analyze how sensitively and consistently different metrics reflect quality changes.

\begin{figure}
\includegraphics[width=\textwidth]{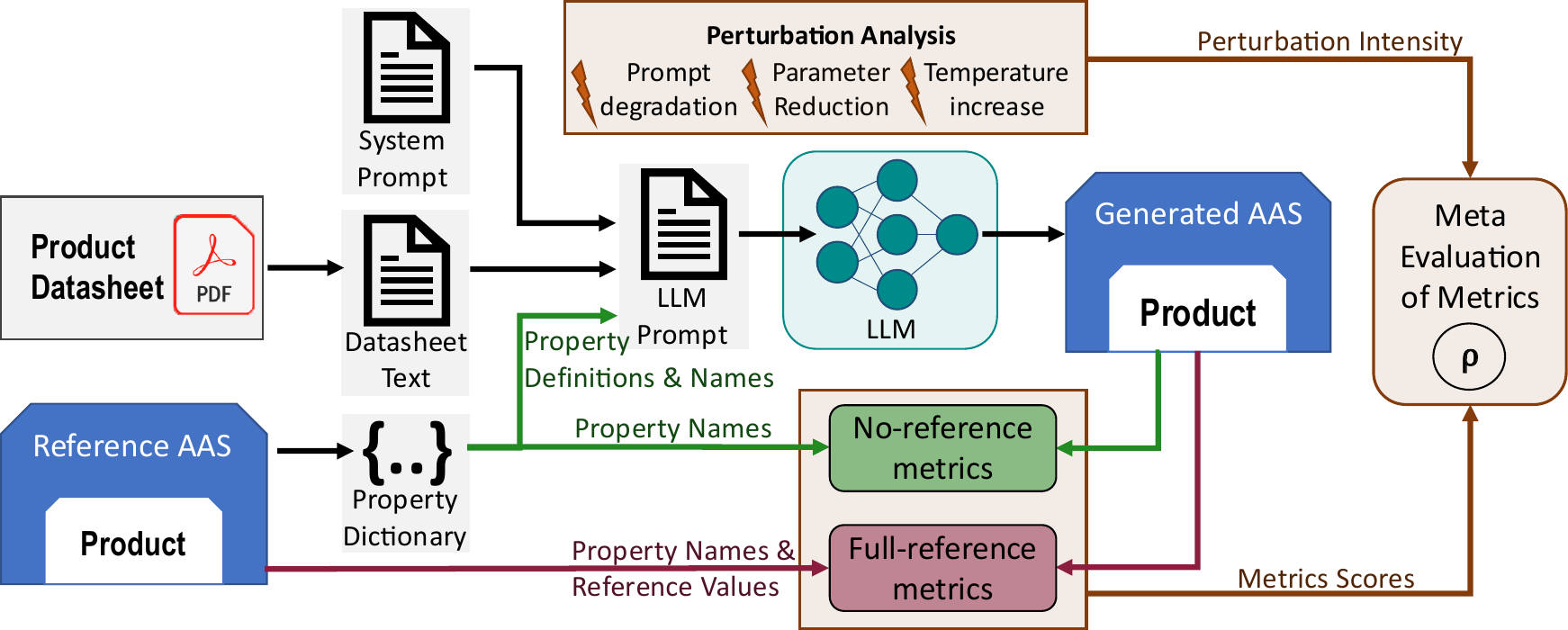}
\caption{Overview of the evaluation pipeline for AAS generation and metric-based quality assessment.} 
\label{fig1}
\end{figure}

The overall evaluation pipeline is illustrated in Fig.~\ref{fig1}. Starting from product datasheets and corresponding AAS, we construct structured prompts that combine the unstructured datasheet content with a predefined property dictionary derived from the reference AAS. This dictionary specifies the expected properties, including their names, value types, and units, and serves as a schema to guide the extraction process.

Based on these prompts, AAS instances are generated using LLM-based information extraction. The model is instructed to extract values for the specified properties and return the results in a structured format, which is then transformed into a technical data submodel of the AAS.

To simulate varying quality levels, controlled perturbations are introduced at different stages of the generation process, including prompt degradation, temperature variation, and the use of different model sizes. These perturbations systematically affect the completeness and correctness of the extracted properties.

The resulting AAS instances are evaluated using both full-reference and no-reference metrics. Full-reference metrics compare generated properties and values against the reference AAS, while no-reference metrics assess intrinsic plausibility, such as alignment with the expected property schema and the presence of missing values.

Finally, a meta-evaluation step analyzes the relationship between perturbation intensity and metric scores. By correlating metric values with controlled degradations, we assess how sensitively and consistently each metric reflects changes in AAS quality.

The following subsections describe the evaluated metrics, perturbation mechanisms, dataset, and experimental setup.

\subsection{AAS Quality Metrics}
We define a set of AAS-specific quality metrics for evaluating AI-generated AAS. These metrics focus on measurable differences in technical properties within AAS submodels. Broader aspects such as schema compliance or adherence to AAS templates are not considered, as they can be addressed by deterministic validation tools.

We distinguish between two complementary types of metrics: \textbf{full-reference metrics} and \textbf{no-reference metrics}. Full-reference metrics compare a generated AAS against a reference and assess similarity in structure, properties, and values. While interpretable and reliable, such references are often unavailable in practice. No-reference metrics instead assess plausibility based on intrinsic characteristics, such as alignment with expected property names or missing values, making them suitable for large-scale quality assurance. Both types are therefore complementary.

To match generated properties to reference properties, we use a two-step procedure. First, similarity scores are computed for all property pairs using either normalized Levenshtein distance or cosine similarity of embeddings. Second, the Hungarian (linear sum assignment) algorithm determines an optimal one-to-one matching, ensuring that each generated property is matched at most once. A similarity threshold is applied to classify matches, optimized via grid search to maximize correlation with perturbation intensity.

For full-reference metrics, matched property values are additionally compared. Numeric values must fall within a tolerance threshold, while string values are evaluated using similarity thresholds.

\begin{table}
\caption{Overview of the evaluated metrics dimensions: reference type, matching type, and IE metric.}\label{tab1}
\begin{tabular}{p{0.13\textwidth} >{\raggedright}p{0.39\textwidth} p{0.46\textwidth}}
Dimension       & Characteristics                       & Description \\
\hline \hline

\multirow{2}{0.13\textwidth}{Reference Type} & \textbf{Name} (no reference values required)   & \multirow{2}{0.46\textwidth}{Whether metric compares property names (against prompt) or property values (against reference AAS).} \\
        & \textbf{Value} (requires reference values) & \\ 
\hline
\multirow{2}{0.13\textwidth}{Matching Type}   & \textbf{Exact} (string equality)               & \multirow{3}{0.46\textwidth}{How property names are matched. For similarity-based matching, pairs match if similarity exceeds a threshold $\tau$. Numeric values match if their relative difference is below 1\%.} \\
                & \textbf{Embedding-based} (cosine similarity) & \\
                & \textbf{String-based} (normalized-Levenshtein-distance-based similarity) & \\
\hline
IE Metric       & \textbf{Precision} & \multirow{3}{0.46\textwidth}{Standard information extraction metrics computed after the matching.} \\ 
                & \textbf{Recall} & \\
                & \textbf{F1 Score} & \\
\end{tabular}
\end{table}

Overall, we evaluate 18 metrics covering all combinations of three dimensions: (i) reference type (names vs. values), (ii) matching type (exact, embedding-based, string-based), and (iii) information extraction metric (precision, recall, F1). Table~\ref{tab1} summarizes these dimensions.
These metrics provide a simple yet effective basis for evaluating structured outputs, enabling both controlled benchmarking and scalable quality checks.

\subsection{Perturbation-Based Evaluation Approach}
To evaluate metric effectiveness, we introduce controlled perturbations during AAS generation to simulate varying quality levels. Metric sensitivity is then assessed by measuring the rank correlation (Spearman’s $\rho$) between perturbation intensity and metric scores.
We consider three perturbation types:

\paragraph{LLM Temperature Variation}
The temperature parameter controls output randomness. Lower values produce deterministic outputs, while higher values increase diversity but typically degrade performance in logic tasks \cite{Renze.2024}. We exploit this effect to vary output quality.

\paragraph{Model Size Reduction}
Model size is reduced using smaller or distilled variants. While the functionality is largely preserved, performance degradation can be observed \cite{Kaplan.2020}, enabling controlled quality variation.

\paragraph{Prompt Degradation}
The input prompt plays a central role in guiding the LLM and is therefore a potential leverage point for introducing controlled perturbations \cite{Moradi.2021}. To analyze metric sensitivity, we apply a set of complementary perturbations at varying intensity levels (0.0–1.0). These perturbations introduce both syntactic and semantic distortions while preserving the overall task intent. The following perturbation types are used:
\begin{itemize}
    \item \texttt{Semantic Drift:} Up to 10\% of tokens are replaced with contextually plausible synonyms (WordNet \cite{Miller.1995}), introducing subtle shifts in meaning.
    
    \item \texttt{Grammar Degradation:} A portion of grammatical elements such as determiners, adpositions, auxiliary verbs, and punctuation are removed based on part-of-speech tagging, reducing syntactic clarity.
    
    \item \texttt{Information Ambiguity:} Hedge phrases (e.g., “maybe,” “or something”) are inserted at random positions (up to 10\% of tokens), increasing ambiguity.
    
    \item \texttt{Contextual Irrelevance:} Grammatically correct but semantically unrelated sentences are inserted after up to 10\% of sentences, introducing distracting context.
    
    \item \texttt{Lexical Noise:} Up to 10\% of characters are modified (substitution, deletion, insertion, transposition) to simulate typographical errors.
\end{itemize}

All perturbations are scaled according to the specified intensity and applied jointly, resulting in a gradual transition from clean to heavily degraded prompts.

An example of prompt perturbation at maximum intensity (1.0) is shown in Table~\ref{tab2}. The perturbed version is obtained by sequentially applying all perturbation types, resulting in a heavily distorted but still partially interpretable prompt. Despite severe degradation, LLMs often recover the task intent, highlighting both their robustness and the challenge of evaluating quality.

\begin{table}
\caption{Original and perturbed prompt at maximum perturbation intensity (1.0).}\label{tab2}
\begin{tabular}{p{0.49\textwidth} | p{0.49\textwidth}}
Original Prompt       & Perturbed Prompt (Intensity = 1.0)\\
\hline \hline
You act as a text API to extract technical properties from a given datasheet. The datasheet will be surrounded by triple backticks (```).  & 
You cat textula matter API to estrat you know technical a liottle bti 
ropejrtes given datasheet datasheet surrounded tyripebackticks 
\end{tabular}
\end{table}

\subsection{Dataset and Preprocessing}
The evaluation is based on 200 products from four manufacturers (A–D), with 50 products each. Products were selected to maximize product diversity, while excluding incomplete or unusable data. Most products belong to the ECLASS segments “27: Electrical engineering” and “51: Fluid Power.” Table~\ref{tab3} summarizes dataset statistics.

AASX files were preprocessed to ensure compatibility with the basyx-python-sdk. This included normalizing deprecated URIs, correcting file references, merging duplicates, standardizing decimal formats, and sanitizing identifiers.
Products were excluded if they used AAS version less than 3, lacked technical data submodels, or had unreadable datasheets. Additionally, only products with at least 10 purely technical properties not related to company or product identification were included.

To account for company-specific structures, a custom property dictionary was derived for each AAS, specifying property names, definitions, value types, and units. Table~\ref{tab4} summarizes the configurations.

The code to run the experiments and links to the product data is publicly available \footnote{https://github.com/janek-gross/experiments}.

\begin{table}
\caption{Data statistics. Number of technical properties per product and product coverage shown by the number of ECLASS categories (including broad product segments and increasingly detailed product groups, classes, and subclasses).}\label{tab3}
\begin{tabular}{p{1.5cm} p{1.4cm} >{\raggedright}p{2.7cm} p{1.4cm} p{1.4cm} p{1.4cm} p{1.4cm}}
Company & Product count & Average property count (Std.\ Dev.) & Segment count & Group count & Class count & Subclass count \\
\hline \hline
A & 50 & 35.4 (12.3) & 1 & 1 & 1 & 5 \\
B & 50 & 36.8 (4.8) & 2 & 5 & 6 & 18 \\
C & 50 & 86.4 (33.9) & 1 & 5 & 6 & 8 \\
D & 50 & 23.9 (9.7) & 3 & 10 & 21 & 30 \\
\hline
Overall & 200 & 45.6 (30.6) & 4 & 17 & 33 & 60
\end{tabular}
\end{table}

\begin{table}
\caption{Fields used in the creation of custom property dictionaries for different manufacturers.}\label{tab4}
\begin{tabular}{p{1.3cm} | p{2.4cm} | p{1.7cm} | p{1.7cm} | p{0.9cm} | >{\raggedright}p{2.2cm} | p{1.2cm}}
Company & Property Name & Definition & Value Type & Unit & Selection Constraints & ECLASS Release \\
\hline \hline
A & id\_short & description & string & None & \textgreater15 properties & 11.0 \\
\hline
B & display\_name & concept\_ description. description  & value\_type & unit* & ECLASS ids not ending in 90-99 \newline \textgreater 32 properties & 12.0 \\
\hline
C & preferred\_name* & definition* & value\_type & unit* & - & 12.0 \\
\hline
D & preferred\_name* & definition* & value\_type & unit* & \textgreater{}9 properties & 14.0 \\
\hline
\multicolumn{7}{l}{*from the data\_specification\_content in the ConceptDescription.}
\end{tabular}
\end{table}

\subsection{Models and Experimental Procedure}

We evaluate both proprietary and open-source LLMs. Proprietary experiments use gpt-4o-mini via the OpenAI API, while open-source evaluations use Qwen3 (0.6B–32B) and DeepSeek-R1 (1.7B–70B) via a self-hosted Ollama deployment. All experiments were conducted on GPU-enabled nodes of an HPC cluster.

The pipeline constructs prompts from datasheets and property dictionaries, instructing the model to extract property records. A json schema was used to structure the LLM output as a list of records with \textit{name}, \textit{value}, \textit{unit} and text \textit{reference} for each record, ensuring machine-readable results.

Generated AAS are evaluated using the defined metrics, and perturbations are applied to analyze metric sensitivity under controlled conditions.


\section{Results}
\label{results}

We first report data statistics to provide an overview of the generation results. In total, we generated 6,400 technical data submodels across 32 experimental conditions, covering different perturbation levels, model variants, and companies. Overall, 231,276 properties were extracted from 200 product datasheets. 

\begin{figure}
\centering
\includegraphics[width=0.87\textwidth]{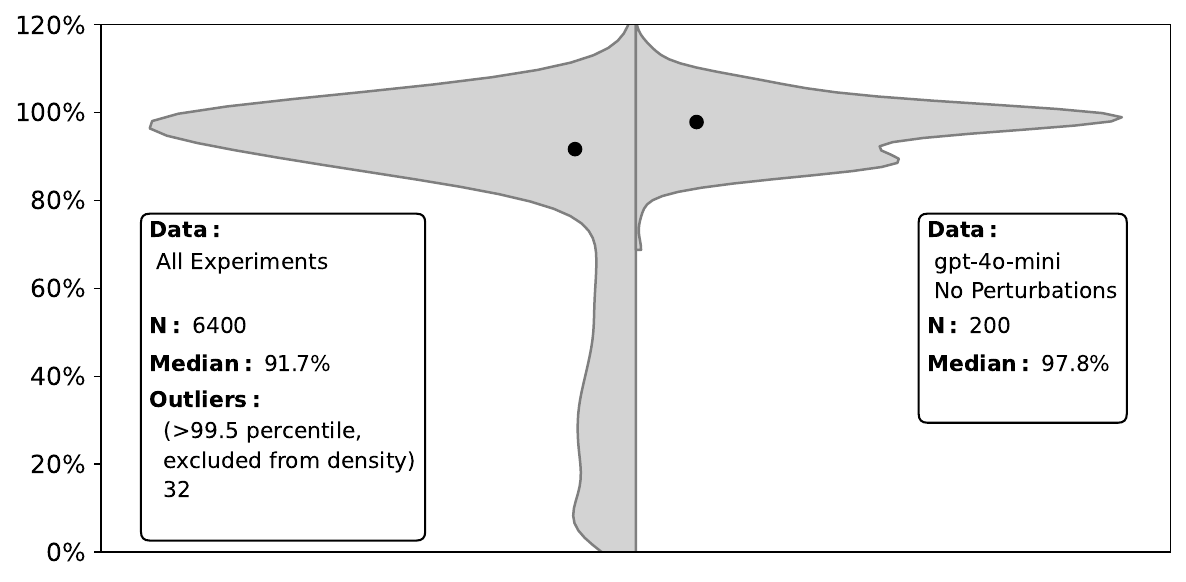}
\caption{Distribution of the ratio of extracted to prompted properties. Values larger than 100\% indicate LLM-hallucinations or duplicate extractions. Values smaller than 100\% indicate incomplete extractions.} 
\label{fig3}
\end{figure}

On average, 83.5\% of the prompted properties were extracted (correct or false) per submodel. In a baseline scenario without perturbations, gpt-4o-mini extracted values for 96.8\% of prompted properties. Fig.~\ref{fig3} shows the distribution of the ratio of extracted to prompted properties. The left density includes all perturbations, while the right density represents an unperturbed baseline scenario.

\begin{figure}
\includegraphics[width=\textwidth]{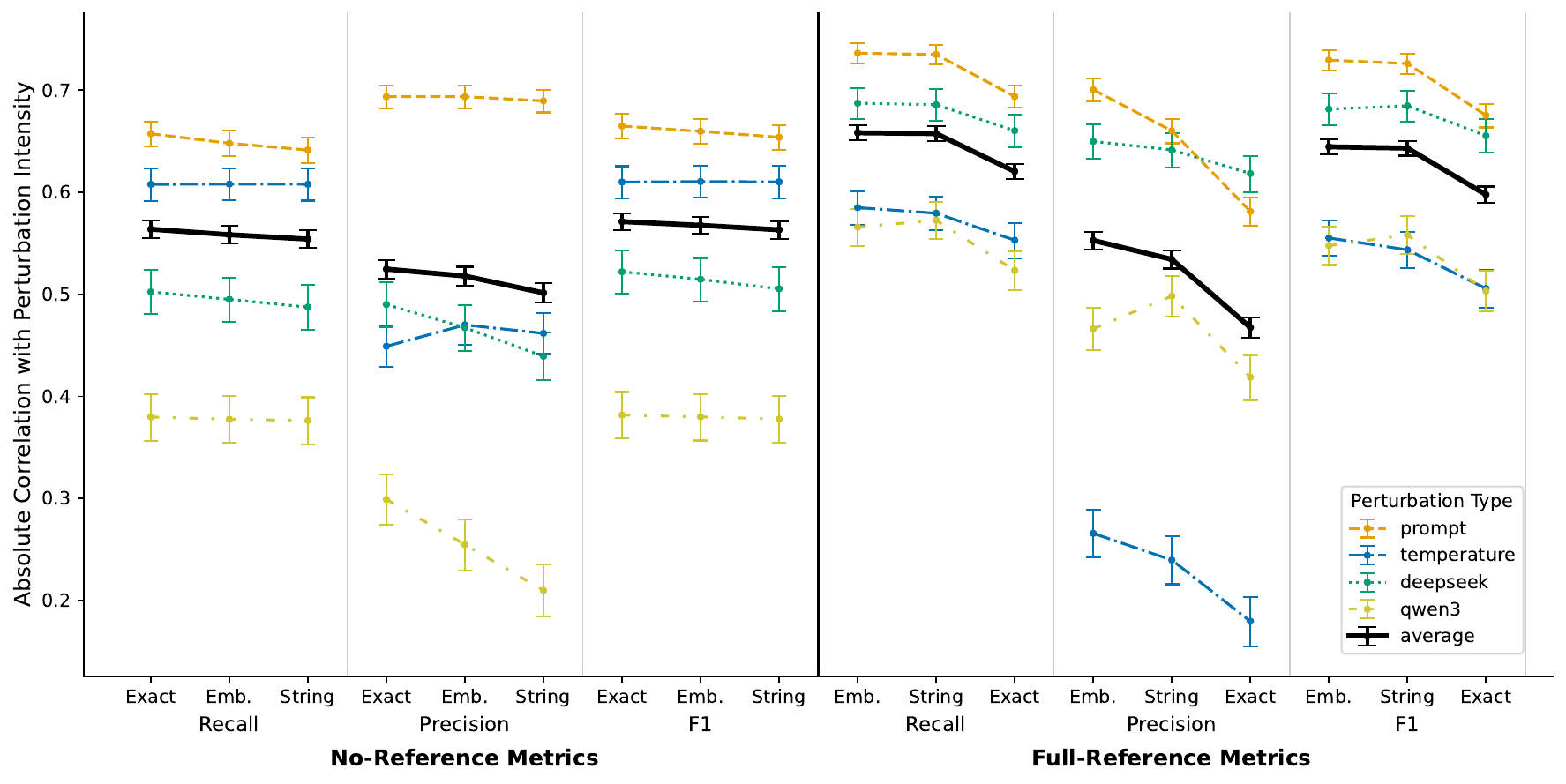}
\caption{Metric sensitivity measured as Spearman’s rank correlation with perturbation intensity. Error bars indicate the standard error of the correlation.} 
\label{fig4}
\end{figure}

\subsection{Metric Performance}

We evaluate metric performance by analyzing the correlation between metric scores and perturbation intensity. Metric performance is assessed using Spearman’s rank correlation between metric values and the intensity of controlled perturbations (temperature, model size, and prompt degradation), addressing RQ1 and RQ2. 

The results are shown in Fig.~\ref{fig4}, which reports the absolute value of the Spearman correlation between metric scores and perturbation intensity for all combinations of reference type, matching strategy and information extraction metric (see Table~\ref{tab1}). Higher values indicate greater sensitivity of a metric to quality degradation. 

Several consistent trends emerge. In the no-reference setting, embedding-based matching performed better than string-based matching. However, neither provided a clear advantage over simple exact matching. For illustration purposes in Fig.~\ref{fig4}, we therefore apply a consistent matching threshold of $\tau = 0.88$ to both string- and embedding-based no-reference metrics even though these metrics would otherwise default to exact matching with an optimal matching threshold very close or equal to $\tau = 1.0$. \textbf{No-reference F1} scores exhibited the strongest monotonic relationships with perturbation intensity.

We utilize these observations and fix name-matching to exact matching in the full-reference setting to isolate the effect of value matching. In this scenario, soft matching based on embeddings resulted in the highest correlations. Among all candidates, \textbf{full-reference recall}, where cosine similarity was applied to values of exactly matching property names achieved the highest average correlation ($\rho = 0.65$) at a value matching threshold of $\tau = 0.88$.
In both settings, precision-based metrics showed limited sensitivity to quality variations. 


\subsection{Impact of Perturbations}

\begin{figure}
\includegraphics[width=\textwidth]{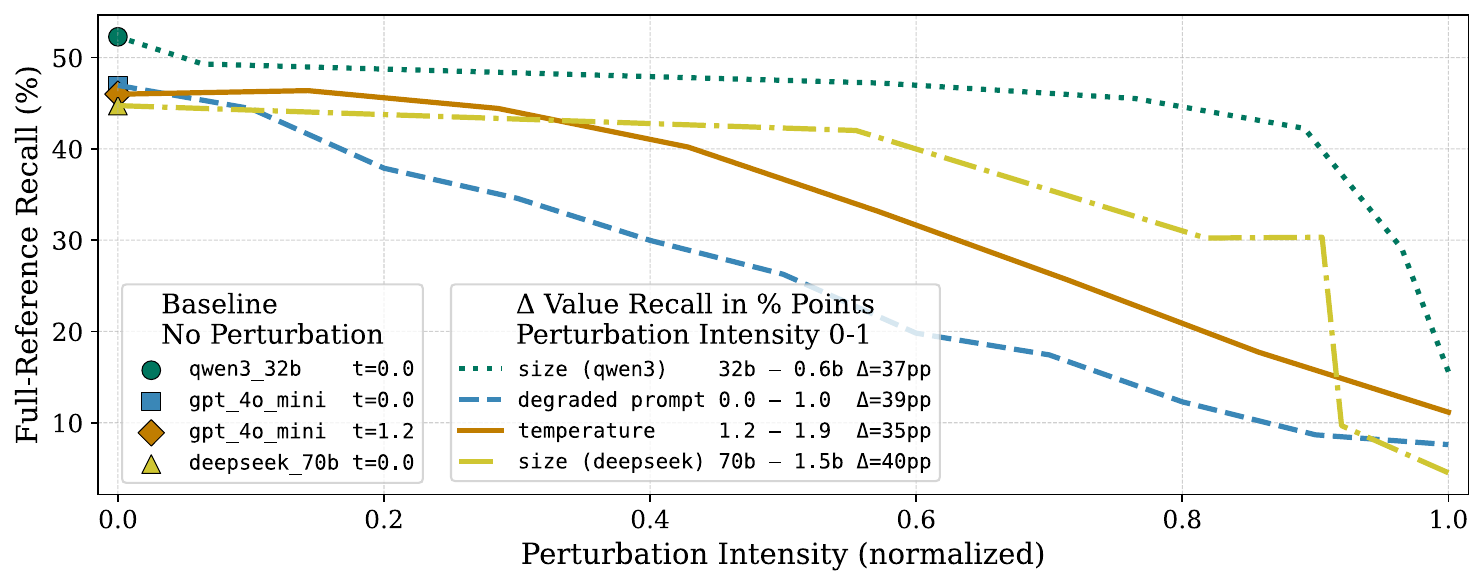}
\caption{Effect of perturbation types on value recall.} 
\label{fig5}
\end{figure}

We next analyze the impact of perturbation types and model configurations on AAS generation quality, addressing RQ2. Fig.~\ref{fig5} illustrates the effects of different perturbations on the most sensitive metric full-reference recall, with perturbation intensity normalized between 0 and 1. Baseline performance for each model is indicated by markers on the left.

Across all perturbations, performance decreases by 35–40 percentage points, confirming that all perturbation types substantially degrade AAS quality. The effect of model size is distinctly non-linear: performance declines gradually for large models but drops sharply below approximately 8 billion parameters, suggesting a non-linear, potentially logarithmic relationship between model size and quality. This behavior further justifies the use of Spearman’s rank correlation, which captures monotonic but non-linear relationships. 

In contrast, both temperature increases and prompt degradation show a more linear degradation trend across the tested ranges. Interestingly, in the baseline scenario, the open-source model Qwen3:32b achieves the best performance, indicating that architecture and training method can outweigh high parameter count alone.

\subsection{Manufacturer and Product Segment Analysis}

\begin{figure}
\includegraphics[width=\textwidth]{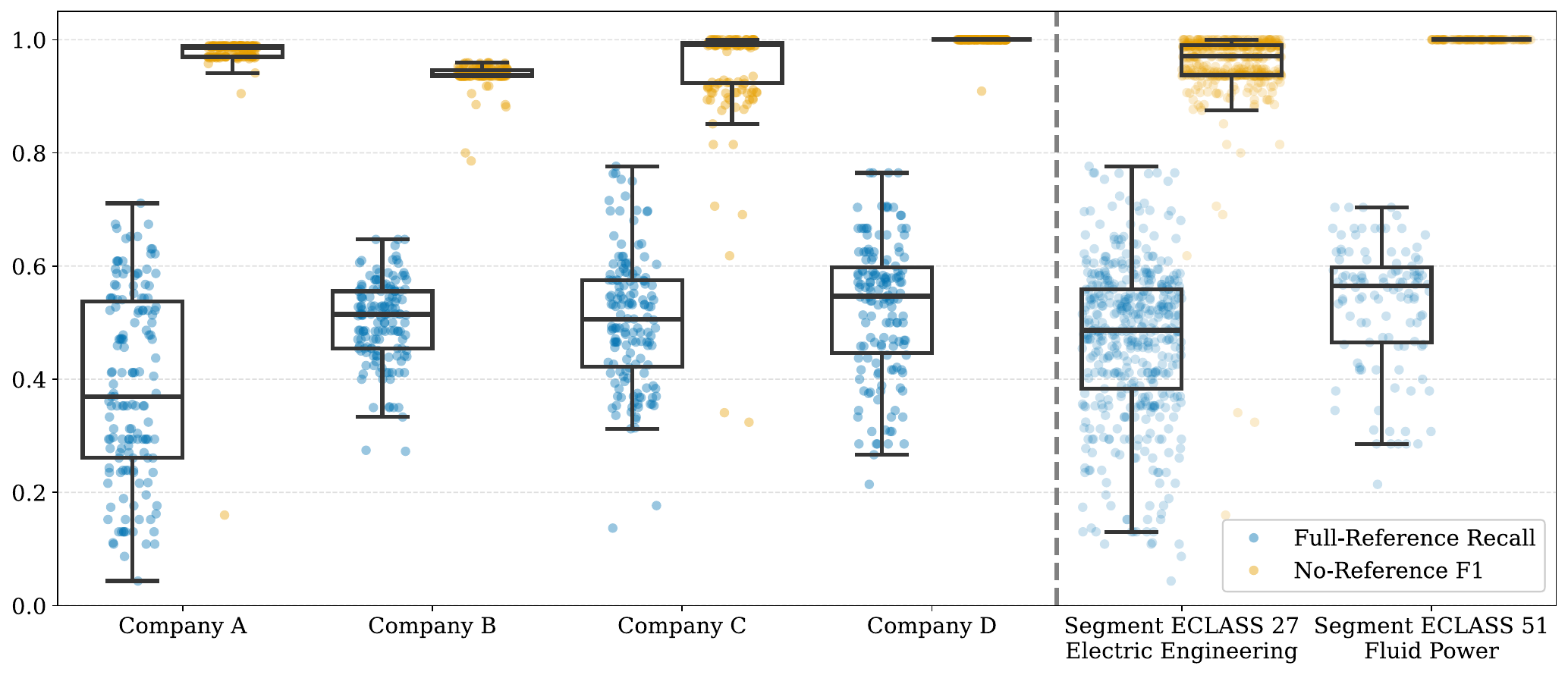}
\caption{AAS quality across manufacturers and product segments.} 
\label{fig6}
\end{figure}

Finally, we analyze how AAS quality varies across manufacturers and product segments (RQ3). Fig.~\ref{fig6} compares both dimensions using strip plots overlaid with boxplots. The left plot shows differences between manufacturers, while the right plot highlights variation across product segments.

ANOVA indicates statistically significant differences between manufacturers for both no-reference and full-reference metrics ($p < 0.001$), with effect sizes of $\eta^2 \approx 0.13$–$0.17$, suggesting that company affiliation explains a moderate portion of the observed variance. This indicates that product-specific characteristics have a stronger influence on AAS quality than manufacturer-specific factors.

At the segment level, Fluid Power products tend to achieve higher scores than Electrical Engineering products, although substantial variability remains within each segment. For no-reference metrics, a ceiling effect is observed, with many samples receiving near-perfect scores, reflecting that only indirect quality signals are captured in the absence of reference values.

\section{Discussion}
\label{discussion}

The results provide several insights into the evaluation of AI-generated AAS and the behavior of different quality metrics.
A key finding is that for matching property names, simple, interpretable metrics based on exact matching outperform more complex similarity-based approaches. This indicates that exact matching more reliably captures quality degradation—such as missing or incorrectly named properties—while embedding-based approaches are intended to capture semantic similarity, they introduce noise in settings where precise and standardized terminology is required. However, for the comparison of property values exact matching turns out too restrictive. Here, a soft matching based on similarity which permits spelling variations and semantically equivalent phrasing leads to the most sensitive metrics.

From a practical perspective, these findings are directly relevant for Industry 4.0 applications. In industrial pipelines, quality assessment must be transparent, reproducible, and easy to integrate. The identified metrics meet these requirements and can be used for benchmarking AAS generation tools, monitoring extraction quality, and supporting acceptance decisions prior to system integration. In particular, no-reference metrics enable scalable quality assurance in scenarios where reference AAS are unavailable.

At the same time, the results highlight limitations of metric-based evaluation. Optimizing a system with respect to a single metric can lead to metric overfitting, where models exploit metric-specific weaknesses rather than improving actual output quality. This is particularly relevant for no-reference metrics, which capture quality only indirectly and may exhibit ceiling effects, as observed in our experiments. In practice, robust evaluation therefore requires a combination of complementary metrics and, where necessary, human-in-the-loop validation.

The perturbation-based methodology itself provides additional insights. By systematically degrading inputs and analyzing metric responses, it enables controlled and reproducible assessment of metric sensitivity without requiring labeled ground truth data. This approach is not limited to AAS generation and can be applied to other structured information extraction tasks and LLM-based systems with end-to-end evaluation requirements.

Finally, differences across model families, manufacturers, and product segments indicate that AAS generation quality depends on both model characteristics and data properties. While larger models tend to perform more robustly, the results also show that architecture and training can outweigh parameter count. Variation between product segments further suggests that domain-specific factors, such as terminology consistency and the number and complexity of properties, significantly influence extraction performance. These findings highlight the importance of context-aware evaluation and caution against relying on single benchmark scenarios.

\section{Threats to Validity}
\label{threats}

Our evaluation is subject to several threats to validity.
\paragraph{Construct Validity}
A central threat concerns the operationalization of AAS quality. In this work, quality is primarily measured through the correctness and completeness of extracted property names and values in technical data submodels. Although these measurements are important, they do not cover all aspects of AAS quality, such as appropriateness in a specific industrial context, compliance with domain-specific modeling conventions, or usefulness for downstream applications. In particular, proposed no-reference metrics measure plausibility only indirectly and fail to detect semantic errors.

\paragraph{Internal Validity}
Our perturbation-based methodology assumes that increasing perturbation intensity corresponds to decreasing AAS quality. Although this assumption is well motivated for the considered perturbation types, such as prompt degradation, temperature increase, and model size reduction, the relationship may not always be strictly monotonic for all models and products. Furthermore, the optimization of similarity thresholds based on correlation with perturbation intensity may bias the evaluation in favor of metrics that align particularly well with our perturbation design.

\paragraph{Conclusion Validity}
The conclusions drawn from the statistical analyses may be affected by variability in product characteristics, model behavior, and perturbation effects. While the dataset and number of generated AAS are substantial, some subgroup analyses, for example by manufacturer or product segment, are based on smaller effective sample sizes. Furthermore, Spearman's rank correlation only measures monotonic relationships, but overlooks non-monotonic effects.

\paragraph{External Validity}
The generalizability of our findings is limited by the scope of the dataset and the evaluated models. Our experiments are based on 200 products from four manufacturers and focus on technical data submodels, with strong representation from specific ECLASS segments. The results may therefore not directly transfer to other industrial domains, other types of AAS submodels, or different product documentation styles. In addition, only a selected set of proprietary and open-source LLMs was evaluated, so the findings may not fully generalize to other model families or future model generations.

\section{Conclusion and Future Work}
\label{conclusion}

In this work, we presented a systematic empirical framework for evaluating the quality of AI-generated Asset Administration Shells (AAS). By combining full-reference and no-reference metrics with a perturbation-based evaluation methodology, we enabled a controlled and reproducible analysis of how different metrics reflect variations in AAS quality.
Our results show that metrics based on exact matching of names and similarity-based matching of values, provide the most reliable indicators of quality degradation across different perturbation types. These findings support the use of soft-matching techniques for benchmarking AAS generation tools in case of non-unique spelling and phrasing of property values.

Beyond the specific results, the proposed perturbation-based methodology provides a general approach for evaluating metrics in scenarios where labeled ground truth is scarce or unavailable. It enables systematic comparison of metrics under controlled conditions and can be transferred to other structured information extraction tasks and LLM-based systems.

In future work, we plan to extend this approach in several directions. First, we aim to incorporate expert assessments to better align automated metrics with human judgment and to calibrate metrics in realistic evaluation settings. Second, we will investigate more advanced evaluation methods, including semantics-aware and LLM-based metrics, and analyze their sensitivity under controlled perturbations. Third, we plan to extend the study to additional AAS submodels and broader industrial domains to assess the generalizability of our findings. Finally, we aim to integrate the proposed metrics into adaptive evaluation pipelines that support continuous quality monitoring and improvement of AI-assisted AAS generation systems.
\begin{credits}
\subsubsection{\ackname}
This work was supported by the research training group “Dependable AI Assistants for the Management of Dynamic Production Systems and Supply Chains (VAMoS)” at Mainz University of Applied Sciences and the Rheinland-Palatinate Technical University of Kaiserslautern-Landau, funded by the Ministry of Science and Health of Rhineland-Palatinate. The authors gratefully acknowledge this support.
\subsubsection{\discintname}
The authors have no competing interests to declare that are
relevant to the content of this article.
    
\end{credits}

%
%
%
\bibliographystyle{splncs04}
\bibliography{references}

\end{document}